%% file: main.tex
\documentclass[conference]{IEEEtran}
\IEEEoverridecommandlockouts
\usepackage{cite}
\usepackage{amsmath,amssymb,amsfonts}
\usepackage{algorithmic}
\usepackage{graphicx}
\usepackage{textcomp}
\usepackage[table,xcdraw]{xcolor}
\usepackage{hhline}
\usepackage{enumitem}
\usepackage[caption=false]{subfig}

\input{preamble.tex}

\def\BibTeX{{\rm B\kern-.05em{\sc i\kern-.025em b}\kern-.08em
    T\kern-.1667em\lower.7ex\hbox{E}\kern-.125emX}}
\begin{document}

\title{\huge{Distributed-Memory Sparse Kernels for Machine Learning}
 \\
}

\author{
    \IEEEauthorblockN{Vivek Bharadwaj\IEEEauthorrefmark{1}, Ayd\i n Bulu\c{c}\IEEEauthorrefmark{1}\IEEEauthorrefmark{2}, James Demmel\IEEEauthorrefmark{1}}
    \IEEEauthorblockA{\IEEEauthorrefmark{1}\textit{EECS Department, University of California, Berkeley, USA}}
    \IEEEauthorblockA{\IEEEauthorrefmark{2}\textit{Computational Research Division, Lawrence Berkeley 
    National Laboratory, Berkeley, USA}}
}

\maketitle

\begin{abstract}
Sampled Dense Times Dense Matrix Multiplication (SDDMM) and Sparse Times Dense Matrix Multiplication (SpMM) appear in diverse settings, such as collaborative filtering, document clustering, and graph embedding. Frequently, the SDDMM output becomes the input sparse matrix for a subsequent SpMM operation. Existing work has  focused on shared memory parallelization of these primitives. While there has been extensive analysis of communication-minimizing distributed 1.5D algorithms for SpMM, no such analysis exists for SDDMM or the
back-to-back sequence of SDDMM and SpMM, termed FusedMM.  
We show that distributed memory 1.5D and 2.5D algorithms for SpMM can be converted to algorithms for SDDMM with identical communication costs and input / output data layouts. Further, we give two communication-eliding strategies to reduce costs further for FusedMM kernels: either reusing the replication of an input dense matrix for the SDDMM and SpMM in sequence, or fusing the local SDDMM
and SpMM kernels. 

We benchmark FusedMM algorithms on Cori, a Cray 
XC40 at LBNL, using \erdosrenyi\ random matrices and large real-world sparse matrices. On 256 nodes with 68 cores each, 1.5D FusedMM algorithms using 
either communication eliding approach can save at least 30\% of time spent exclusively in communication compared to executing a distributed-memory SpMM and SDDMM 
kernel in sequence. Our 2.5D communication-eliding algorithms save 21\% of communication time compared to the unoptimized sequence. On real-world matrices
with hundreds of millions of edges, all of our algorithms exhibit at least a
10x speedup over the SpMM algorithm in \petsc. On these matrices, our 
communication-eliding techniques exhibit runtimes up to 1.6 times faster 
than an unoptimized sequence of SDDMM and SpMM. 
We embed and test the scaling of our algorithms in real-world applications, including collaborative filtering via alternating-least-squares and inference for attention-based graph neural networks.
 
\end{abstract}


\section{Introduction}
Sampled Dense-Dense Matrix Multiplication (SDDMM) and Sparse-Times-Dense Matrix Multiplication (SpMM) have
become workhorse kernels in a variety of computations. Their use in matrix completion \cite{canny_big_nodate}
and document similarity computation \cite{tithi_efficient_2021} is well documented, and they are 
the main primitives used in graph learning \cite{velickovic_graph_2018}. 
The recent interest in Graph Neural Networks (GNNs) with self-attention has led libraries such
as Amazon Deep Graph Library (DGL) \cite{wang2019dgl} to 
expose SDDMM  and SpMM primitives to users. Typical applications make 
a call to an SDDMM operation and feed the sparse output to an SpMM operation, repeating the pair several
times with the same nonzero pattern (but possibly different values) for the sparse matrix. We refer 
to the back-to-back sequence of an SDDMM and SpMM as FusedMM.

Several works optimize SDDMM and SpMM kernels in shared memory environments, or on
accelerators such as GPUs~\cite{hong_adaptive_2019,nisa_sampled_2018, jiang_novel_2020}. Separately, there have been prior efforts
~\cite{koanantakool_communication-avoiding_2016,selvitopi_distributed-memory_2021,acer_improving_2016} to optimize distributed memory SpMM by minimizing processor to processor
communication. There is no significant work, however, on 
distributed algorithms for SDDMM or FusedMM. 

We make three main contributions. First, we show that every sparsity-agnostic 
distributed-memory algorithm for SpMM can be converted into a distributed memory 
algorithm for SDDMM that uses the same input / output
data distribution and has identical communication cost. Second, we give two methods to elide 
communication when executing an SDDMM and SpMM in sequence (FusedMM): replication reuse, which elides 
a second replication of an input matrix, and local kernel fusion, which 
allows local SDDMM and SpMM operations to execute on the same processor
without intervening communication.
These methods not only eliminate unnecessary communication rounds, they 
also enable algorithms that replicate dense  matrices to scale to higher or lower replication factors (depending on which 
of the two methods used). Third, we demonstrate, both in theory and practice, these algorithms that
replicate or shift sparse matrices perform best when the ratio of nonzeros in the sparse matrix
to the number of nonzeros in the dense matrices is sufficiently low. When the number of nonzeros
in the sparse matrix approaches the number of nonzeros in either of the dense matrices, algorithms that shift or replicate
dense matrices become favorable. Our work gives, to the best of our knowledge, the first
benchmark of 1.5D SpMM and SDDMM algorithms that cyclically shift sparse matrices and replicate a dense
matrix, which enables efficient communication scaling 
when the input dense matrices are tall and skinny. It also
gives the first head-to-head comparison between sparse shifting 
and dense shifting 1.5D algorithms, which we show
outperform each other depending on the problem setting.

\section{Definitions}

\begin{table}[]
\centering
\begin{tabular}{|c|l|}
\hline
\multicolumn{1}{|l|}{Symbol} & Description                      \\ 
    \hhline{|=|=|}
$S, R$                                  & $m \times n$ sparse matrix       \\
$A$                                     & $m \times r$ dense matrix        \\
$B$                                     & $n \times r$ dense matrix        \\
$\phi$                                  & The ratio $\textrm{nnz(S)} / nr$ \\
$p$                                     & Total processor count            \\ 
$c$                                     & Replication Factor               \\ \hline
\end{tabular}
\vspace{3pt}
\caption{Symbol Definitions}
\vspace{-13pt}
\end{table}

Let $A \in \RR^{m \times r}, B \in \RR^{n \times r}$ be dense matrices and let $S \in \RR^{m \times n}$
be a sparse matrix. The SDDMM operation produces a sparse matrix with sparsity structure identical to $S$
given by
\begin{equation}
\textrm{SDDMM}(A, B, S) := S * (A \cdot B^T)
\label{eq:sddmm}
\end{equation}
where $*$ denotes elementwise multiplication and $\cdot$ denotes matrix-matrix multiplication. Computing the output with a dense matrix multiplication 
followed by element-wise multiplication with $S$ is inefficient, since
we only need to compute the output entries at the nonzero locations of $S$.

For clarity of notation when describing our distributed memory algorithms, we distinguish between the SpMM operation involving $S$ that takes $B$ as an input vs. the 
operation involving $S^T$ that takes $A$ as an input. 
Specifically, define
\begin{align*}
\textrm{SpMMA(S, B)} &:= S \cdot B \\
\textrm{SpMMB(S, A)} &:= S^T \cdot A
\end{align*}

\noindent
where the suffix A or B on each operation refers to the matrix with the same 
shape as the output. The distinction is useful for applications
such as alternating least squares and graph attention networks, which 
require both operations at different points in time. Finally, we borrow 
notation from prior works on the SDDMM-SpMM sequence \cite{rahman_fusedmm_2021} 
and use FusedMMA, FusedMMB to denote operations that are compositions of SDDMM with SpMMA or SpMMB, given as
\begin{align*}
\textrm{FusedMMA}(S, A, B) &:= \textrm{SpMMA}(\textrm{SDDMM}(A, B, S), B) \\
\textrm{FusedMMB}(S, A, B) &:= \textrm{SpMMB}(\textrm{SDDMM}(A, B, S), A)
\end{align*}

\section{Existing Work}

\subsection{Shared Memory Optimization}
Local SpMM and SDDMM operations are bound by memory bandwidth compared to dense matrix
multiplication. Accelerating either SpMM or SDDMM in a shared memory environment, such as a single CPU node or GPU, typically involves blocking a 
loop over the nonzeros of $S$ to optimize cache reuse of the dense matrices~\cite{nisa_sampled_2018}. 
For blocked SDDMM and SpMM kernels, the traffic between fast and slow memory is exactly
modeled by the edgecut-1 metric of a hypergraph partition induced on $S$
(treating the rows of $S$ as hyperedges and the columns as vertices, with nonzeros
indicating pins). 
Jiang et al. \cite{jiang_novel_2020} reorder the sparse matrix to minimize the connectivity metric,
thereby reducing memory traffic. Instead of reordering $S$, Hong et al. 
\cite{hong_adaptive_2019} adaptively choose a blocking 
shape tuned to the
sparsity structure to optimize performance. Both optimizations require 
expensive processing steps
on the sparse matrix $S$, which are typically amortized away by repeated calls to the kernel. 

\subsection{Distributed Sparsity-Aware Algorithms}
We can divide distributed-memory implementations for both SpMM and SDDMM into two categories: sparsity-aware
algorithms, and sparsity-agnostic bulk communication approaches. In the former
category, the dense input matrices are partitioned among processors along with the sparse
matrix nonzeros $(i, j)$, such that each processor owns at least one of $A_{i:}$ or $B_{j:}$.
When processing $(i, j)$, if a processor does not own one of the two dense rows needed, it fetches
the embedding from another processor that owns it~\cite{acer_improving_2016}. 
Such approaches work well when $S$ is very sparse. 
They also benefit from graph /
hypergraph partitioning to reorder the sparse matrix, which can reduce the number of remote fetches 
that each processor must make while maintaining load balance. Such approaches suffer, however,
from the overhead of communicating the specific embeddings requested by processors,
which typically requires round-trip communication. As $S$ gets denser,
processors are better off broadcasting all of their embeddings.

\subsection{Sparsity Agnostic Bulk Communication Algorithms}
Sparsity agnostic algorithms resemble distributed 
dense matrix multiplication algorithms by
broadcasting, shifting, and reducing block rows and block columns of $A, B$, 
and $S$. These methods cannot significantly benefit from graph partitioning
and they often rely on a random permutation of the sparse matrix to load balance among processors. 

Such algorithms are typically described as 1D, 1.5D, 2D, 2.5D, or 3D. 1D and 2D 
algorithms are memory-optimal, with processors requiring no 
more aggregate memory than the
storage required for inputs and outputs (up to a small constant factor for
communication buffering). 1.5D, 2.5D, and 3D algorithms increase collective memory
consumption of at least one of the three operands to asymptotically
decrease communication costs. In this work, we will consider only 1.5D and 2.5D
algorithms, since 1D, 2D, and 3D algorithms are special cases of these two.

Koanantakool et al. show when $A, B$ and $S$ are all square, 1.5D SpMM 
algorithms that cyclically shift the sparse matrix yield superior performance 
\cite{koanantakool_communication-avoiding_2016}. 
They only benchmark multiplication of all square matrices, which does not cover 
the more common case where $r \ll m, n$. 
For graph embedding problems, $r$ is typically between $64$ and $512$, 
whereas $m$ and $n$ can be in tens to hundreds 
of milllions. For this case of tall-skinny dense 
matrices, Tripathy et al. 
introduced CAGNET, which trains graph neural networks on hundreds of GPUs using 
1.5D and 2.5D distributed SpMM operations \cite{tripathy_reducing_2020}. They, 
along with  Selvitopi et al. 
\cite{selvitopi_distributed-memory_2021}, showed that 2D algorithms for SpMM 
suffer from diminished arithmetic intensity 
as processor count increases. In contrast, both demonstrate that 1.5D algorithms 
communicating dense matrices exhibit excellent scaling with processor count. Selvitopi et al. did not consider 2.5D algorithms, however, 
and neither work benchmarked 1.5D algorithms that cyclically shift sparse matrices. In addition, the 2.5D algorithms in CAGNET 
only replicate the dense matrix, whereas it is also possible to construct implementations
that only replicate the sparse matrix.

\subsection{Background on Dense Distributed Linear Algebra}
Our sparsity-agnostic algorithms resemble 1.5D and 2.5D variants of the the Cannon and 
SUMMA distributed dense GEMM algorithms \cite{cannon}, \cite{summa}. 
In the 2.5D Cannon-like algorithm to compute 
compute the matrix product $X = YZ$, the submatrix domains of the output $X$ are \textbf{replicated} 
among processors \cite{solomonik_25d}.
E.g., for every entry $X_{ij}$, different processors compute different parts of the
inner product $Y_{i:} \cdot Y_{:j}$ and reduce their results at the end. 
The inputs are both \textbf{propagated} during the algorithm: submatrices of $A$ and $B$ are
cyclically shifted between processors in stages. The SUMMA algorithm replaces the stages
of cyclic shifts with broadcast collectives. 1.5D variants of these algorithms keep at least one of 
the three matrices \textbf{stationary}: submatrices of a stationary matrix are distributed among
processors, but they are not broadcast, reduced, or shifted. It is possible to modify these
algorithms so that an input matrix is replicated rather than an output.

Our algorithm design is dictated by choosing which submatrices 
we keep stationary, replicate, and
propagate. While these choices do not matter for dense GEMM if all matrices are square, they impact our 
kernels due to the sparsity of $S$ and the extremely rectangular shapes of $A$ and $B$.

Kwasniewski et al.\,recently proposed COSMA \cite{cosma}, which uses the classic red-blue 
pebbling game to design an optimal parallelization scheme for distributed dense GEMM with matrices of varying shapes. 
Their communication-minimizing algorithms achieve excellent performance relative to SCALAPACK and 
recent high performance matrix multiplication work. COSMA, however, does 
not account for sparsity in either inputs or outputs. Our work, furthermore, focuses on the
FusedMM computation pattern commonly found in applications, which provides 
further opportunities for communication minimization than considering the SpMM and SDDMM kernels
in isolation. 

\section{Distributed Memory Algorithms for SDDMM and FusedMM}
This section highlights the connection between SpMM and SDDMM and gives a high-level procedure to convert between algorithms that compute each one. This enables us to use a single input distribution to compute SDDMM, SpMMA, and
SpMMB, at the cost of possibly replicating the sparse matrix $S$ by a factor of 2 to store its transpose. Each kernel requires the same amount of communication.
Next, we give two \textit{communication elision} strategies when we 
execute an SDDMM and an SpMM operation in sequence,
with the output of the SDDMM feeding into the SpMM (a FusedMM operation). 
Optimizing for the sequence of the two kernels reduces communication overhead compared to simply executing one 
distributed algorithm followed by the other.
Subsequently, we detail the implementation of those high-level strategies.

\subsection{The Connection between SDDMM and SpMM}
SDDMM and SpMM have identical data access patterns, which becomes clear when we compare their serial algorithms
(take SpMMA here). Letting sparse matrix $R$ be the SDDMM output 
and letting $(i, j)$ index a nonzero entry of $S$,
we have $R_{ij} = S_{ij} \paren{A_{i:} \cdot B_{j, :}^T}$
with $R$ set to zero where $S$ is 0. Compare this equation 
to each step required to compute $A \mathrel{{+}{=}} \textrm{SpMMA}(S, B)$:
for every nonzero $(i, j)$ of $S$, we perform the update $A_{i:}  \mathrel{{+}{=}} S_{ij} B_{j:}$.
For both SDDMM and SpMMA, each nonzero of $S$ 
results in an interaction between a row of $A$ and a row of $B$.

Now consider any distributed memory algorithm for SpMMA that does not replicate its input or output matrices during
computation.
For each nonzero $(i, j)$ and for every index  $k \in \left[1, r\right]$, this algorithm must co-locate $S_{ij}$, $A_{ik}$, and $B_{jk}$ on 
some processor and compute $A_{ik}  \mathrel{{+}{=}} S_{ij} B_{jk}$. Transform the algorithm as follows:

\begin{enumerate}[topsep=0pt,itemsep=-1ex,partopsep=1ex,parsep=1ex,leftmargin=*]
    \item Change the input sparse matrix $S$ to an output matrix initialized
    to 0.
    \item Change $A$ from an output matrix to an input matrix.
    \item Have each processor execute local update $S_{ij} \mathrel{{+}{=}} A_{ik} B_{jk}$.
\end{enumerate}

Then for all nonzeros $(i, j)$ and every index $k \in \left[1, r\right]$, the processors collectively execute computations to
overwrite $S$ with $A B^T$ masked at the nonzeros of $S$. This is exactly the SDDMM operation up to multiplication of $A B^T$
with the values initially in $S$. If the output distribution of $S$ is identical to its
input distribution after execution, then the post-multiplication with the initial values in $S$ does not require additional communication. A similar transformation converts an algorithm for SpMMB into an algorithm for SDDMM. 

We can extend this transformation procedure to algorithms that replicate input and output matrices. Typically, inputs are replicated via broadcast, while output replication requires a reduction at the end
of the algorithm to sum up temporary buffers across processors. Since we interchange in the input / output role between matrices $A$
and $S$, we convert broadcasts of the values of $S$ in SpMMA to reductions of its values in SDDMM, and reductions of $A$
to broadcasts. 

Algorithms \ref{alg:15d_shift} and \ref{alg:25d_dense_replicating} illustrate the transformation procedure outlined above 
by giving unified
algorithms to compute SDDMM, SpMMA, and SpMMB. The local update executed by each processor changes based on the kernel
being executed, and initial broadcasts / terminal reductions depend on whether matrices function as inputs or outputs. 

\subsection{Strategies for Distributed Memory FusedMM}

\begin{figure*}
  \centering
  \includegraphics[trim={1.25cm 12.5cm 2cm 1.25cm}, scale=0.40, clip]{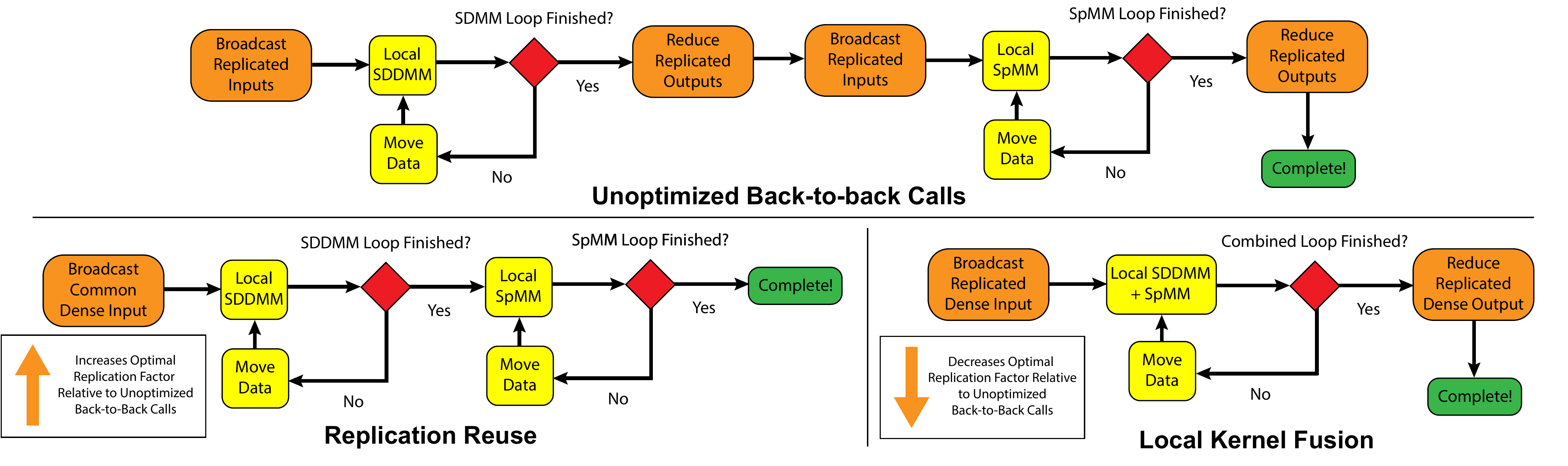}
  \caption{
   Communication eliding strategies illustrated. Crucially, the benefit from replication reuse and local kernel fusion arises from an increased or decreased optimal replication factor, 
   not just the elimination of communication phases. 
}

  \label{fig:fusion_flowchart}
  \vspace{-12pt}
\end{figure*}

FusedMMA computes each row of its output as
\begin{equation*}
\textrm{FusedMMA}(A, B, S)_{i:} := \sum_{j\ \vert\ (i, j) \in nnz(S)} S_{ij} \gen{A_{i:}, B_{j:}} B_{j:}
\end{equation*}
which is a sum of rows $j$ of matrix $B$ weighted by the dot products between rows $j$ of $B$ and rows
$i$ of $A$. The analogous equation for FusedMMA replaces $S_{ij}$ 
with $S^T_{ij} = S_{ji}$. The simplest distributed implementation for FusedMMA 
computes the intermediate SDDMM, stores it temporarily,
and feeds the result directly to SpMMA, exploiting the common data layout 
for inputs and outputs explainting in the previous section. 
The communication cost for this implementation is twice that of
either a single SpMMA operation or SDDMM operation. 
Such an algorithm for FusedMMA takes the following structure: 

\begin{enumerate}[topsep=0pt,itemsep=-1ex,partopsep=1ex,parsep=1ex,leftmargin=*]
    \item Replicate dense matrices $A, B$ in preparation for SDDMM (if either $A$ or $B$ is replicated) 
    \item Propagate matrices, compute SDDMM
    \item Reduce SDDMM output $R$ (if sparse matrix is replicated) 
    \item Replicate dense matrix $B$ in preparation for SpMMA (if $B$ is replicated)
    \item Propagate matrices, compute SpMMA
    \item Reduce output matrix $A$ (if $A$ is replicated)
\end{enumerate}
In the outline above, propagation refers to any communication that excludes replication of inputs / outputs (for example, cyclic shifts
of buffers in Cannon's algorithm). Increased replication factors allow us to decrease propagation cost, but increase the required
memory and result in a higher cost to perform the replication. Choosing the optimal replication factor involves minimizing the sum
of the communication overhead paid in replication and propagation. We can save communication in two ways: first by reducing the replication cost, and second by reducing the propagation cost. Figure
\ref{fig:fusion_flowchart} illustrates both. 

\noindent
\textbf{(1) Replication Reuse: } This approach 
    performs only a single replication of an input matrix in both the SDDMM and SpMM computations. If we replicate the dense matrix $B$, before the SDDMM operation,
    we don't require a second replication before the SpMM phase, nor do we need a reduction of the output buffer. 

\noindent
\textbf{(2) Local Kernel Fusion: } 
This approach combines the two propagation steps, 2 and 5, into a single phase while only replicating
    matrix $A$. Using locally available data at each propagation step, 
    it requires performing a local SDDMM and SpMM in sequence without intermediate communication. 
    Thus, local kernel fusion with any data distribution
    that divides $A$ and $B$ by columns would yield an incorrect result. From the standard definition of 
    FusedMMA, we must compute the dot product 
    $\gen{A_{i:}, B_{j:}}$ that scales every row of $B$ before aggregating those rows, requiring us to complete the
    SDDMM before performing any aggregation. The 1.5D algorithm that replicates and shifts dense matrices 
    (Section~\ref{sec:algdesc}) co-locates entire rows of $A$ and $B$ on the same processor during the stages of computation, 
    and is the only candidate that can take advantage of local kernel
    fusion. Besides the communication savings that they offer, optimized local
    FusedMM functions (e.g., \cite{rahman_fusedmm_2021}) 
    can improve performance by eliding intermediate
    storage of the SDDMM result.

Although applying either method would provide moderate communication savings
without modifying any other aspect of the algorithm, their utility really
lies in allowing us to change the degree of replication for 
any algorithm that replicates a dense matrix.
Algorithms that employ replication reuse can achieve lower communication overhead at \textit{higher} replication factors compared to
an unoptimized sequence of calls. Intuitively, increasing the 
replication factor 
enables the algorithm to trade away more overhead in the 
propagation phase before
the cost of replication becomes overwhelming. By contrast, algorithms that 
employ local kernel fusion can 
achieve lower communication overhead at \textit{lower} replication factors. 
The algorithm requires less replication
to balance off the lower cost of propagation. Note that
these strategies are mutually exclusive; applying them
both to 1.5D dense shifting algorithms would require
propagating a separate accumulation buffer in the propagation
phase, which destroys the benefit of local kernel fusion.

While we have discussed strategies for FusedMMA, we obtain algorithms for FusedMMB by interchanging the roles of $A$ and $B$ 
and replacing matrix $S$ with its transpose $S^T$. In practice, this amounts to storing two copies of the sparse matrix across
all processors, one with the coordinates transposed.

\section{Algorithm Descriptions}
\label{sec:algdesc}
\begin{figure}
    \centering
    \includegraphics[scale=0.70, trim={30cm 28cm 32cm 4.6cm, 
    clip}]{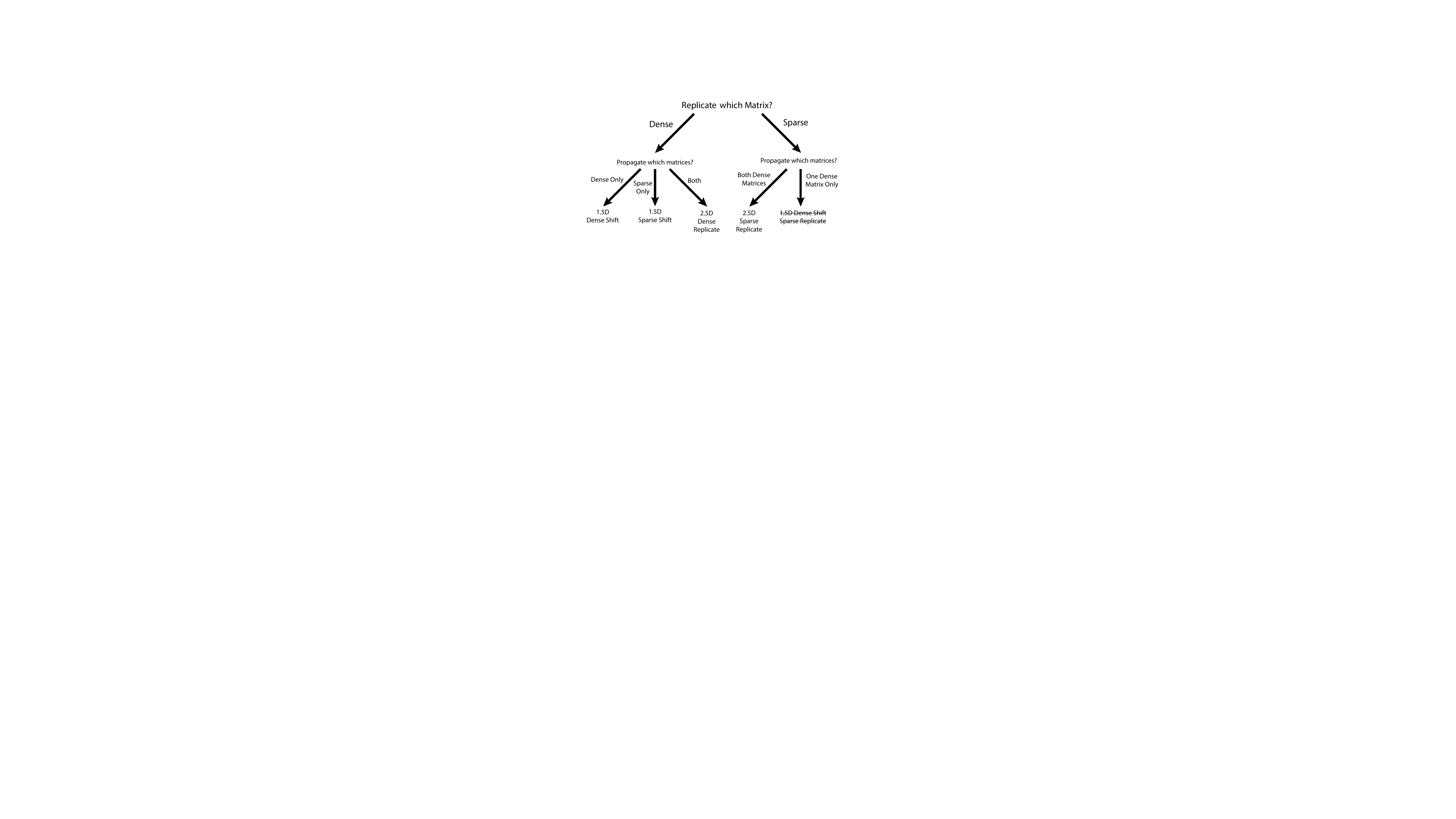}
    \vspace{6pt}
    \caption{Design Choices in SpMM Algorithms. We do not consider the 1.5D sparse replicating, 
    dense shifting algorithm, since it is inferior to the 2.5D sparse replicating algorithm.}
    \label{fig:taxonomy}
    \vspace{-5pt}
\end{figure}

We consider formulations for SpMM detailed previously~\cite{koanantakool_communication-avoiding_2016,tripathy_reducing_2020,selvitopi_distributed-memory_2021}, but some with key modifications. We arrive at each algorithm by deciding which of the three matrices $A, B$, and $S$ to replicate,
propagate, or keep stationary (see figure \ref{fig:taxonomy}). 
We consider formulations where only a single matrix is replicated
enabling us to scale to higher replication factors and take advantage of both communication-eliding strategies
described above. 

\begin{figure*}
  \centering
  \includegraphics[trim={1cm 18.9cm 1cm 1.2cm}, scale=0.95, clip]{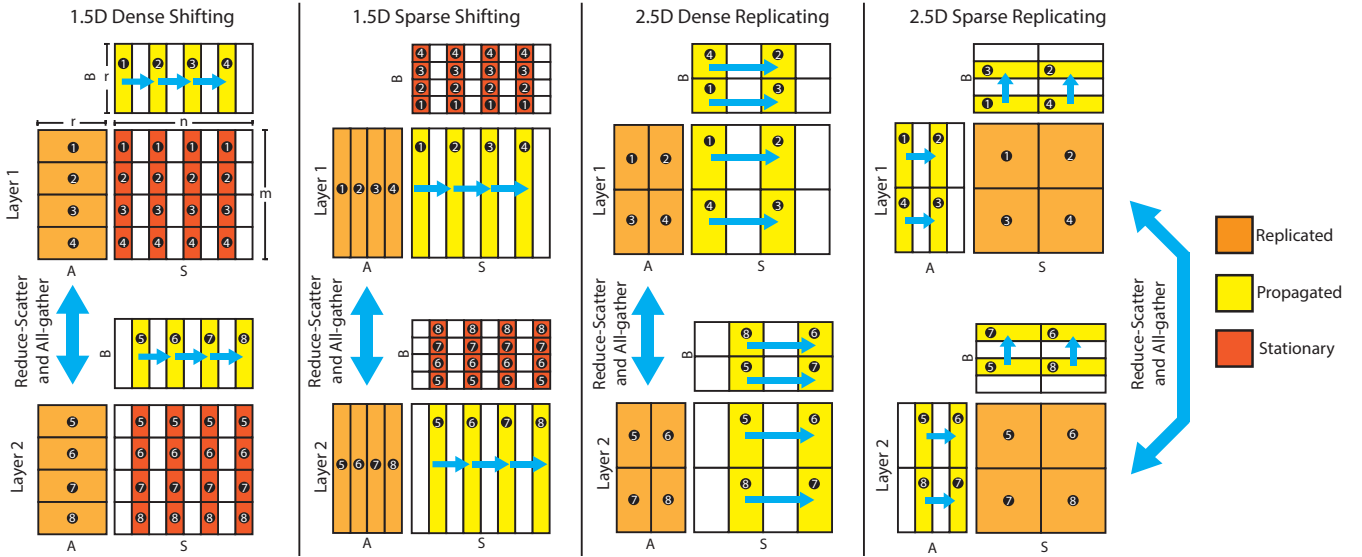}
  \caption{
 Data movement after replication, illustrated for $p=8$ processors and replication factor 
 $c = 2$. Within each layer, submatrices shown in yellow are cyclically shifted from processor 
to processor (propagation) in the direction indicated by single-headed blue arrows. Submatrices 
in orange are replicated across layers and participate 
in reduce-scatter / all-gather operations outside the propagation phase. 
Submatrices in red remain stationary on each processor. 
Black numbered circles give the processor owning each submatrix at the beginning of each algorithm (note the initial skew for 2.5D algorithms). 
}
  \label{fig:data_movement}
  \vspace{-12pt}
\end{figure*}

Let $p$ be the total processor count. We list the input distributions 
of all matrices in Table~\ref{table:input_distributions}. Each matrix is partitioned by
blocks into a grid with the specified dimension. The product of these dimensions 
may exceed the total processor count because processors can own multiple non-contiguous
blocks, as in block row / column cyclic distributions. The third column of Table~\ref{table:input_distributions} 
gives the processor that initially owns a block $(i, j)$ as an 
$(u, v)$ or $(u, v, w)$ tuple, which
identifies the processor position within a 2D / 3D grid. 
For the 2.5D sparse replicating algorithm, ``*" means all
processors along the third axis of the grid share the 
coordinates of block $(i, j)$. Figure \ref{fig:data_movement} illustrates 
the data movement in our algorithms for 8 processors with a replication factor of 2. We
will refer to the grid axis along which inputs are reduced or gathered as the ``fiber
axis". It is the second dimension of the computational grid for 1.5D algorithms and
the third dimension for 2.5D algorithms.

To analyze our algorithms, we use the 
$\alpha{\text -}\beta \text{-}\gamma$ model where
$\alpha$ is the per-message latency, $\beta$ is the inverse-bandwidth, and $\gamma$ is the cost per FLOP performed
locally on each processor. Since our algorithms communicate blocks of matrices, they
exchange at most a small multiple of $p$ messages, each of which is a large block of a 
matrix. Therefore, we focus on minimizing the number of data words communicated by each processor, the coefficient
of inverse-bandwidth in our model. In the analysis that follows, we use ``communication cost" to 
mean the maximum amount of time that any processor spends sending and receiving messages.

We assume sends and receives can make independent progress on each node and use
the costs for collectives in the literature~\cite{chan_collective_2007}. To aid the 
analysis, assume $m \approx n$, and let  $\phi$ be the ratio of the
number of nonzeros in $S$ to the number of nonzeros of 
dense matrix $B$, i.e. $\phi = \textrm{nnz}(S) / nr$. 

\begin{table}[]
\centering
\begin{tabular}{|l|l|l|}
\hline
\multicolumn{1}{|c|}{\textbf{Matrix}} & \multicolumn{1}{c|}{\textbf{Grid}}             & \multicolumn{1}{c|}{\textbf{Owner of Block $(i, j)$}} \\ \hline
\multicolumn{3}{|l|}{\cellcolor[HTML]{C0C0C0}1.5D Dense Shifting}              \\ \hline
A      & $p \times 1$                                   & $(i / c, i \% c)$    \\
B      & $p \times 1$                                   & $(i / c, i \% c)$     \\
S,R    & $(p / c) \times p$                             & $(i, j \% c)$        \\ \hline
\multicolumn{3}{|l|}{\cellcolor[HTML]{C0C0C0}1.5D Sparse Shifting}             \\ \hline
A      & $p \times (p / c)$                                   & $(j, i \% c)$     \\
B      & $p \times (p / c)$                                   & $(j, i \% c)$     \\
S,R    & $1 \times p $                             & $(j / c, j \% c)$             \\ \hline
\multicolumn{3}{|l|}{\cellcolor[HTML]{C0C0C0}2.5D Dense Replicating}           \\ \hline
A      & $\sqrt{pc} \times \sqrt{p / c}$               & $(i / c, j, i\% c)$   \\
B      & $\sqrt{pc} \times \sqrt{p / c}$               & $(i / c, j, i \% c)$                     \\
S,R    & $\sqrt{p/c} \times \sqrt{pc}$                 & $(i, j / c, j \% c)$  \\ \hline
\multicolumn{3}{|l|}{\cellcolor[HTML]{C0C0C0}2.5D Sparse Replicating}          \\ \hline
A      & $p \times \sqrt{p/c}$                  & $(i / c, j, i \% c)$  \\
B      & $\sqrt{p/c} \times p$                  & $(i, j / c, j \% c)$  \\
S,R    & $\sqrt{p/c} \times \sqrt{p/c}$ & $(i, j, *)$          \\ \hline
\end{tabular}
\vspace{2mm}
\caption{Input Matrix Distributions Before Replication}
\label{table:input_distributions}
\vspace{-12pt}
\end{table}

\subsection{1.5D Dense Shifting, Dense Replicating}
1.5D algorithms operate on a $(p/c) \times c$
processor grid, where $c \geq 1$ is the replication factor. We can 
interpret the 1.5D dense shifting, dense replicating 
algorithm as $c$ layers 
of concurrently executing 1D algorithms. To decrease communication as the processor count increases, 
the 1.5D dense shifting, dense replicating algorithm replicates one of the two dense matrix inputs, 
propagates the other dense matrix, and keeps the remaining sparse matrix stationary on each processor. 

The procedure is detailed in Algorithm~\ref{alg:15d_shift} and illustrated in Figure~\ref{fig:data_movement}. 
The sparse matrix $S$ is stored with a column block cyclic distribution across
the grid layers. The processors begin by allocating a buffer $T$ for the replicated matrix $A$, keeping $T$
initially 0 if $A$ is an output buffer and otherwise 
all-gathering blocks of $A$ of size $(n/p) \times r$ 
within each fiber axis if $A$ is an input (replication). 
For $p / c$ phases, algorithms cyclically shift their local blocks
of the matrix $B$ (propagation). Finally, if $A$ is the output of the computation, the buffer $T$ is reduce-scattered
to processors within the fiber. Increasing the replication factor $c$ decreases communication 
costs incurred within each layer due to cyclic shifts of $B$, but 
increases the communication cost of all-gather / reduce-scatter primitives between layers.

\begin{algorithm}
\SetAlgoLined
 \KwIn{Mode $\in \set{\textrm{SDDMM, SpMMA, SpMMB}}$}
 \KwData{Dense Matrices $A$, $B$, and sparse matrices $S, R$ distributed on a 
 $\frac{p}{c} \times c$ grid}
 \KwResult{One of $A$, $B$, or $R$ filled with the output of an SpMMA, SpMMB, or SDDMM computation,
 depending on Mode}
    $T := \textrm{Zeros}(c m / P, R)$\;
    \lIf{Mode $\in \set{\textrm{SDDMM, SpMMB}}$}{ \\
       \Indp Allgather($A_{loc}$, $T$, fiber-axis)
    }
    \Indm

 \For{$i = 1...p/c$}{
    \lIf{Mode == SDDMM}{ \\
        \Indp $R_{loc} \mathrel{{+}{=}} \textrm{SDDMM}(T, B_{loc}, 
        S_{loc, i})$
    }
    \Indm
    \lElseIf{Mode == SpMMA}{ \\
        \Indp $T \mathrel{{+}{=}} \textrm{SpMMA}(S_{loc, i}, B_{loc})$
    }
    \Indm
    \lElse{ \\
        \Indp $B_{loc} \mathrel{{+}{=}} 
        \textrm{SpMMB}(S_{loc, i}, T)$
    }
    \Indm
    Cyclic Shift $B_{loc}$ within layer\; 
 }
 \lIf{Mode == SpMMA}{ \\
     \Indp Reduce-Scatter($T$, $A_{loc}$, fiber-axis)
 }
 \caption{Unified 1.5D Algorithm Moving Dense Matrices for SpMMA, SpMMB, SDDMM}
 \label{alg:15d_shift}
\end{algorithm}
\noindent
\textbf{Communication analysis, No Elision:} Consider a pair of SDDMM and SpMMA operations that execute as two sequential
calls to Algorithm \ref{alg:15d_shift} with no intervening communication elision. 
The all-gather and reduce-scatter primitives operate on dense block 
of size $nrc / p$ within each fiber,
which contains $c$ processors. Since all-gather happens within SDDMM and reduce-scatter happens within SpMMA, 
they communicate $2 ((c-1)/c) (nrc/p)$ words. Each layer contains $p / c$ processors and 
executes $2 p / c$ cyclic shifts of dense blocks with size $nr/p$.
Each cyclic shift 
communicates $nr/p$ words. Multiplying by the number of phases and adding to the cost of communication 
along fibers gives a communication cost 
$ nr (2(c - 1)/p + 2/c).$
Differentiating this expression and setting equal to 0, we find that $c = \sqrt{p}$ minimizes the cost. 

\noindent
\textbf{Communication analysis, Replication Reuse:} 
If we apply replication reuse to optimize FusedMMA (by interchanging the roles of $A$ and $B$,
replacing $S$ with $S^T$, and performing an SpMMB), we eliminate the terminal reduce-scatter operation,
yielding a communication cost $ nr ((c - 1)/p + 2/c).$
The optimal replication factor $c$ becomes $c = \sqrt{2p}$. Since we have less overall communication within
each fiber, we can afford to increase replication further to 
drive down the cost of cyclic shifts within each layer.
The ratio of the communication cost to the version without FusedMM elision for
the optimal choice of $c$ in each is
$(1 - 2 \sqrt{2p})/(2 - 4 \sqrt{p})$,

which for $p \rightarrow \infty$ tends to $1 / \sqrt{2}$. Thus, we save roughly 30\% of
communication compared to executing two kernels in sequence. 

\noindent
\textbf{Communication analysis, Local Kernel Fusion:} 
If we use the local kernel fusion strategy to optimize FusedMMA, we need only a single round 
of $p / c$ cyclic shifts
instead of two. We still require an initial all-gather and terminal reduce-scatter, giving $nr (2(c - 1)/p + 1/c)$ words communicated.
The smaller communication cost from cyclic shifts yields an optimal replication factor $c = \sqrt{p / 2}$, since
communication within each fiber costs more relative to the cyclic shifts.
The ratio of the communication cost of local kernel fusion to the case without
any communication elision for the optimal choice of replication factors approaches 
$1 / \sqrt 2$ as $p \rightarrow \infty$,
meaning that either communication eliding strategy produces the same reduction in communication. Local kernel fusion, however, is the only algorithm that permits
local FusedMMA operations on each processor, since it does not 
divide row and column embeddings along the short $R$-axis. 

\subsection{1.5D Sparse Shifting, Dense Replicating}
1.5D algorithms that shift the sparse matrix operate analogously to the dense shifting, dense
replicating algorithm. In this case, the sparse matrix is propagated while the non-replicated dense
matrix is stationary on each processor. Koanantakool et al.~\cite{koanantakool_communication-avoiding_2016} showed that 1.5D SpMM algorithms
are more efficient than 2.5D algorithms when $m{=}n{=}r$. 
Their work, however, both replicates and shifts the
sparse matrix to decrease latency at high processor counts and improve local SpMM throughput. Their approach was
also motivated by working with all square matrices, which 
means that any dense replication proves prohibitively 
expensive in comparison to the sparse matrix communication. By contrast, our
dense matrices are tall-skinny. We cyclically shift the sparse matrix and replicate the 
dense input matrix to reduce communication at high processor counts. This is advantageous when 
$\textrm{nnz}(S)$ is significantly smaller than $nr$. The $c$ layers of the grid 
still participate in reduce-scatters / all-gathers of an input matrix, 
but within each layer, block rows of the sparse matrix $S$ are cyclically shifted, and we now divide $A$ and $B$ by block columns rather
than block rows. Dividing the dense matrices by columns,
however, can significantly hurt local kernel throughput on some architectures \cite{tripathy_reducing_2020}.

\noindent
\textbf{Communication Analysis, FusedMM}: We again apply replication reuse to optimize the FusedMM operation. 
The algorithm incurs the communication cost of $2p / c$ cyclic
shifts of an average of $\textrm{nnz}(S) / p$ nonzeros each; each nonzero consists of three words 
when the sparse matrix is in coordinate format. We add this to a single all-gather operation on blocks
of size $nr / p$ across the $c$ processes in each fiber to produce an aggregate cost
\begin{equation}
 6 \paren{\frac{\textrm{nnz}(S)}{c}} + \frac{nr(c-1)}{p}
    \label{eq:15d_fused2_bwidth_cost}
\end{equation}
where the first term arises from cyclic shifts and the second term arises from the all-gather.
For ease of analysis, we replace $\textrm{nnz}(S)$ with $\phi nr$ (see the earlier definition of $\phi$) and 
optimize for $c$ to get $c = \sqrt{6p\phi}$. When $\phi$ is low and $c < 1$, we interpret this as indicating 
that no amount of replication is favorable. When $c < 1$, the optimal communication cost is 
$6 \phi n r$, and when $c > 1$ (which occurs at higher processor counts), the communication cost is 
\begin{equation}
\frac{nr}{\sqrt{p}} \paren{2 \sqrt{6\phi} - \frac{1}{\sqrt p}}
    \label{eq:15d_fused2_bwidth_cost_final}
\end{equation}
Ignoring the lower order $1 / \sqrt{p}$ term, 
Equation \ref{eq:15d_fused2_bwidth_cost_final} indicates that
when $\phi$ is low, the sparse shifting algorithm performs better than the 1.5D dense shifting algorithm. As $\phi$
increases, the dense shifting 1.5D algorithm performs better.

\subsection{2.5D Dense Replicating Algorithms}
2.5D algorithms operate on a $\sqrt{p / c} \times \sqrt {p/c} \times c$ grid. We can
interpret each layer as executing a concurrent version of SUMMA / Cannon
on a square 2D grid. This 2.5D algorithm replicates a dense matrix and 
cyclically shifts a sparse matrix and the remaining dense matrix within each layer. Algorithm \ref{alg:25d_dense_replicating} gives pseudocode of the procedure;
processors within each layer cyclically shift both $S$ and $B$ along processor row,
resp. column, axes, while blocks of $A$ are reduce-scattered or all-gathered 
along the fiber-axis at the beginning (and end). 
As written, the algorithm requires an initial shift of its inputs to correctly
index blocks of the matrices. In practice, applications
do not need to perform this initial shift if they fill the input and output
buffers appropriately. Thus, we don't include the initial shift
in our communication analysis.

\noindent
\textbf{Communication Analysis, FusedMM:} We can only apply replication reuse when using 
2.5D dense replicating algorithms, as the input dense matrices are
divided by columns among processors. The communication 
analysis is similar to the 1.5D FusedMM algorithms, so we omit the details
for brevity. The optimal 
replication factor is
$c=p^{1/3}(1 + 3\phi)^{2/3}$. The resulting optimal cost is
\begin{equation*}
        \frac{nr}{p^{2/3}} \paren{
        \frac{2 + 6 \phi}{(1 + 3\phi)^{1/3}} + (1 + 3\phi)^{2/3} - \frac{1}{p^{1/3}}
        } 
        = O \paren{\frac{nr \phi^{2/3}}{p^{2/3}}}
    \label{eq:optimal_bwidth}
\end{equation*}
Notice the factor $p^{2/3}$ in the denominator, as opposed to $p^{1/2}$ for the 1.5D algorithms. As with the 1.5D sparse shifting algorithm, replication is less favorable when $\phi$ is
low and becomes more favorable as $\phi$ increases. 

\begin{algorithm}
\SetAlgoLined
 \KwIn{Mode $\in \set{\textrm{SDDMM, SpMMA, SpMMB}}$}
 \KwData{Dense $A$, $B$, and sparse $S, R$ matrices distributed on a $\sqrt {p/c} 
 \times \sqrt {p/c} \times c$ grid per Table \ref{table:input_distributions}}
 \KwResult{One of $A$, $B$, or $R$ filled with the output of an SpMMA, SpMMB, or SDDMM computation,
 depending on Mode}

 \lIf{Input Matrices are not shifted}{ \\
    \Indp Cyclic Shift $S_{loc}$ to processor \textit{row-rank} - \textit{col-rank}\; 
    Cyclic Shift $B_{loc}$ to processor \textit{col-rank} - \textit{row-rank}
 }        
 \Indm 
 $T := \textrm{Zeros}(c m / P, R)$\;
 \lIf{Mode $\in \set{\textrm{SDDMM, SpMMB}}$}{ \\
    \Indp Allgather($A_{loc}$, $T$, fiber-axis)
 }        
 \Indm  

 \For{$i = 1...\sqrt {p / c}$}{
    \lIf{Mode == SDDMM}{ \\
        \Indp $R_{loc} \mathrel{{+}{=}} \textrm{SDDMM}(T, B_{loc}, 
        S_{loc})$
    }
    \Indm
    \lElseIf{Mode == SpMMA}{ \\
        \Indp $T \mathrel{{+}{=}} \textrm{SpMMA}(S_{loc, i}, B_{loc})$
    }
    \Indm
    \lElse{ \\
        \Indp $B_{loc} \mathrel{{+}{=}} \textrm{SpMMB}(S_{loc, i}, 
        T)$
    }
    \Indm
    Cyclic Shift $S_{loc}$ by 1 within row clockwise\; 
    Cyclic Shift $B_{loc}$ by 1 within column clockwise\; 
 }
 
\lIf{Mode == SpMMA}{ \\
\Indp Reduce-Scatter($T$, $A_{loc}$, fiber-axis)
}
 
\caption{Unified 2.5D Dense Replicating Algorithm for SpMMA, SpMMB, SDDMM}
\label{alg:25d_dense_replicating}
\end{algorithm}
\subsection{2.5D Sparse Replicating Algorithms}
2.5D sparse replicating algorithms operate similarly to the dense replicating version, except that both dense matrices cyclically shift within each layer and the nonzeros
of the sparse matrix are reduce-scattered / all-gathered along the fiber axis. 
The algorithm has the attractive property that only the nonzero values need
to be communicated along the fiber axis, since the nonzero coordinates
do not change between function calls. In contrast to the dense replicating algorithm,
the 2.5D sparse replicating algorithm divides the dense embedding matrices into successively more
block columns as $c$ increases. Because this algorithm does not replicate 
dense matrices, it cannot benefit from communication elision when performing a FusedMM operation. 

\noindent
\textbf{Communication Analysis, No Communication Elision } To execute an SDDMM and SpMM in sequence, the
sparse replicating 2.5D algorithm executes an initial all-gather to accumulate the sparse
matrix values at each layer of the processor grid, and an all-reduce (reduce-scatter + all-gather)
between the SDDMM and SpMM calls. Over both propagation steps, it executes $4\sqrt{p}{c}$ cyclic
shifts of dense blocks containing $nr / p$ words. The optimal 
replication factor is $c = p^{1/3} \paren{2/(3\phi)}^{2/3}$, and 
the resulting optimal communication cost is
\begin{equation*}
        \frac{nr \phi^{1/3}}{p^{2/3}}  \paren{\sqrt[3]{2^5 3 \phi} + \sqrt[3]{2^2 3} - \frac{3\phi^{2/3}}{p^{1/3}}} 
        = O \paren{\frac{nr \phi^{1/3}}{p^{2/3}}} 
    \label{eq:sparse25dbwidthoptimal}
\end{equation*}
Note the factor $\phi^{1/3}$ instead of $\phi^{2/3}$; for $\phi > 1$, this is an improvement,
and when $\phi < 1$ but is sufficiently far away from 0, the difference is subsumed by
the constant factors in front of the expression. Note also that the optimal value of $c$ has 
$\phi^{2/3}$ in its denominator, indicating that a sparser input $S$ benefits from higher replication.

\subsection{Summary}
Table \ref{table:comm_costs} summarizes the analysis in the previous sections by giving communication 
and latency costs for each of the algorithms above embedded in the FusedMM procedure. 
It also lists the dimensions of the matrices used in each local call to either SDDMM or SpMM. 
Table \ref{table:optimal_replication_factors} gives the optimal replication factors for our algorithms. 

Our theory predicts that 1.5D algorithms with correctly tuned
replication factors will marginally outperform the 2.5D algorithms over a range 
of processor counts, sparse matrix densities, and dense matrix widths. 
The choice to use a dense shifting or sparse shifting 1.5D algorithm 
depends on the value of $\phi$ in the specific problem instance. Figure
\ref{fig:optimal_algorithm_selection} (Section~\ref{sec:exp}) 
illustrates and evaluates these predictions.

\begin{table*}[h]
  \centering
  \begin{tabular}{c|c|c|c|c}
    Algorithm & Message Count & Words Communicated & Local $S$-matrix Dim. & Local $B$-matrix Dim. 
    \\
    \hhline{=|=|=|=|=}
    1.5D Dense Shift, Repl. Reuse & $\frac{2p}{c} + (c-1)$ & $nr \paren{\frac{2}{c} 
    + \frac{(c - 1)}{p}}$
    & $\frac{nc}{p} \times \frac{n}{p}$ & $\frac{n}{p} \times r$ \\
    \hline 
    1.5D Dense Shift, Local Kernel Fusion & $\frac{p}{c} + 2(c-1)$ & $nr \paren{\frac{1}{c} + \frac{2(c -1)}{p}}$
    & $\frac{nc}{p} \times \frac{n}{p}$ & $\frac{n}{p} \times r$ \\
    \hline 
    1.5D Sparse Shift, Repl. Reuse & $\frac{2p}{c} + (c - 1)$ & $nr\paren{\frac{6\phi}{c} + \frac{c - 1}{p}}$
    & $\frac{nc}{p} \times n$ & $n \times \frac{r}{p}$ \\ \hline
    2.5D Dense Replicate, Repl. Reuse & $4 \sqrt{\frac{p}{c}} + (c-1)$ 
    &$\frac{nr}{\sqrt{pc}} \paren{6\phi + 2 + \frac{c^{3/2}}{\sqrt p} - \frac{\sqrt c}{\sqrt p}}$ 
    & $\frac{n \sqrt c}{\sqrt p} \times \frac{n}{\sqrt{pc}}$ 
    & $\frac{n}{\sqrt{pc}} \times \frac{r \sqrt{c}}{\sqrt{p}}$ \\ \hline
    2.5D Sparse Replicate, No Elision & $4 \sqrt{\frac{p}{c}} + 3(c - 1)$ &
    $\frac{nr}{\sqrt{p}} \paren{ \frac{4}{\sqrt c} + \frac{3 \phi (c-1)}{\sqrt p} }$
    & $\frac{n \sqrt{c}}{\sqrt{p}} \times \frac{n \sqrt{c}}{\sqrt p}$ &
    $\frac{n}{\sqrt {pc}} \times \frac{r \sqrt c}{\sqrt p}$ 
    \vspace{2mm}
  \end{tabular}
  \caption{Latency, Bandwidth, and Matrix Dimensions in Local Kernel Calls 
  for FusedMM Algorithms}
  \label{table:comm_costs}
  \vspace{-12pt}
\end{table*}
\begin{center}

\begin{table}[]
\begin{tabular}{|l|c|}
\hline
\textbf{Algorithm}                            & \textbf{Best Replication Factor}    \\ \hline
1.5D Dense Shift, No Elision & $\sqrt{p}$                          \\
1.5D Dense Shift, Replication Reuse           & $\sqrt{2p}$                \\
1.5D Dense Shift, Local Kernel Fusion              & $\sqrt{p/2}$                \\
1.5D Sparse Shift, Replication Reuse          & $\sqrt{6p \phi}$                    \\
2.5D Dense Replicate, No Elision & $\sqrt[3]{p\frac{(1 + 3 \phi)^2}{4}}$                      \\ 
2.5D Dense Replicate, Replication Reuse       & $\sqrt[3]{p(1 + 3 \phi)^2}$                      \\
2.5D Sparse Replicate, No Elision & $\sqrt[3]{\frac{p}{ (2 \phi/ 3) ^2}}$                      \\ \hline
\end{tabular}
\vspace{2mm}
\caption{Optimal Replication Factors for FusedMM Algorithms}
\label{table:optimal_replication_factors}
\vspace{-15pt}
\end{table}
\end{center}

\section{Experiments}
\label{sec:exp}
We ran experiments on Cori, a Cray XC40 system at Lawrence
Berkeley National Laboratory, on 256 Xeon Phi Knights Landing (KNL) CPU nodes. 
Each KNL node is single socket CPU containing 68 cores running at 1.4 GHz with access 
to 96 GiB of RAM~\cite{noauthor_cori_nodate}. 
KNL nodes communicate through an Aries interconnect 
with a Dragonfly topology.

Our implementation employs a hybrid OpenMP / MPI programming model, 
with a single MPI rank and 68 OpenMP threads per node.
We use the \verb|MPI_ISend| and \verb|MPI_Irecv| primitives for point-to-point 
communication, as well as the blocking 
collectives  \verb|MPI_Reduce_scatter| and \verb|MPI_Allgather|. 
To load balance among the processors, we randomly permute the 
rows and columns of sparse matrices that we read in. 

We use the Intel Math Kernel Library (MKL version 18.0.1.163) to perform 
local SpMM computations. Because the MKL sparse BLAS does not yet include an SDDMM function, we wrote a simple implementation that uses OpenMP to parallelize the collection of independent dot products required in the computation. We rely on 
CombBLAS \cite{azad_combinatorial_2021} for sparse matrix IO and 
to generate distributed \erdosrenyi\ random sparse
matrices. We use Eigen \cite{guennebaud_eigen_2010} as a wrapper around matrix buffers to handle local 
dense linear algebra. Our code is available online\footnote{https://github.com/PASSIONLab/distributed\_sddmm}.

\subsection{Baseline Comparisons}
Our work presents, to the best of our knowledge, the first distributed-memory implementation of SDDMM for general 
matrices. There is no comparable library to establish a baseline. Among the \petsc, Trillinos, and libSRB 
libraries, only \petsc\ offers a distributed-memory SpMM implementation as a special case of the 
\verb|MatMatMult| routine \cite{petsc-user-ref}, which we compare against. Since \petsc\ does not support hybrid OpenMP / MPI parallelism 
\cite{petsc-threads}, we ran benchmarks with 68 MPI ranks per node (1 per core). 
To ensure a fair benchmark against FusedMM algorithms that make a call to both SDDMM 
and SpMM, we compare our algorithms against two back-to-back SpMM calls from the \petsc\ library. 
Since SDDMM and SpMM have identical FLOP counts and communication requirements,
using two back-to-back SpMM calls offers a reasonable performance surrogate for FusedMM. 

Figure \ref{fig:strongscale} compares the strong scaling performance of our algorithms 
to Cray \petsc\ (v3.13.3.0, 64-bit). The library requires 
a 1D block row distribution for all matrices and does not perform any replication,
resulting in poor communication scaling. Due to exceptionally
poor performance from \petsc\ on sparse matrices with many nonzeros, we omit the baseline
benchmark on the two larger strong scaling workloads.

\subsection{Weak Scaling on \erdosrenyi\ Random Matrices}

\begin{figure*}
    \centering
    \includegraphics[scale=0.50, trim={0cm 0.8cm 0cm 0cm, clip}]{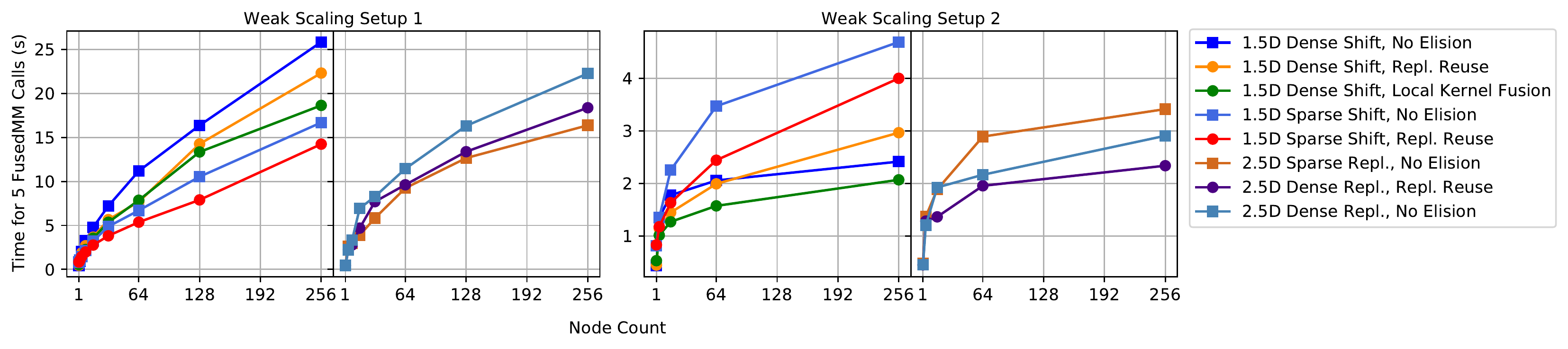}
    \vspace{0pt}
    \caption{Weak scaling experiments with $r=256$. 
    The horizontal axis gives the node count $p$ for
    each experiment. In setup 1, $p$ processors run FusedMM on $S$
    of side-length $2^{16} p$ with 32 nonzeros per row. In setup 2,
    $p$ processors execute FusedMM on $S$ of side length $2^{16} p^{1/2}$
    with $32 p^{1/2}$ nonzeros per row. }
    \label{fig:weakscale}
    \vspace{-12pt}
\end{figure*}

To benchmark the weak scaling of our algorithms, we keep the FLOP count assigned to each node 
constant (assuming a load-balanced sparse matrix) while increasing both the processor count 
and the ``problem size". We investigated two methods of doubling the problem size, 
each giving different performance characteristics.

\noindent
\textbf{Setup 1:} In these experiments, node counts double from experiment to 
experiment as we double the side-length of the 
sparse matrix, keeping the number of nonzeros in each row and the embedding dimension $r$ constant. 
We begin with sparse matrix dimensions $65536 \times 65536$ on a single node with 32 nonzeros per row, and we use
$r=256$ as the embedding dimension. We scale to 256 nodes that collectively process
a $2^{24} \times 2^{24}$ matrix with 500 million nonzeros. 

\begin{figure}
  \centering
  \includegraphics[trim={0cm 1cm 0cm 0cm, clip}, scale=0.375]{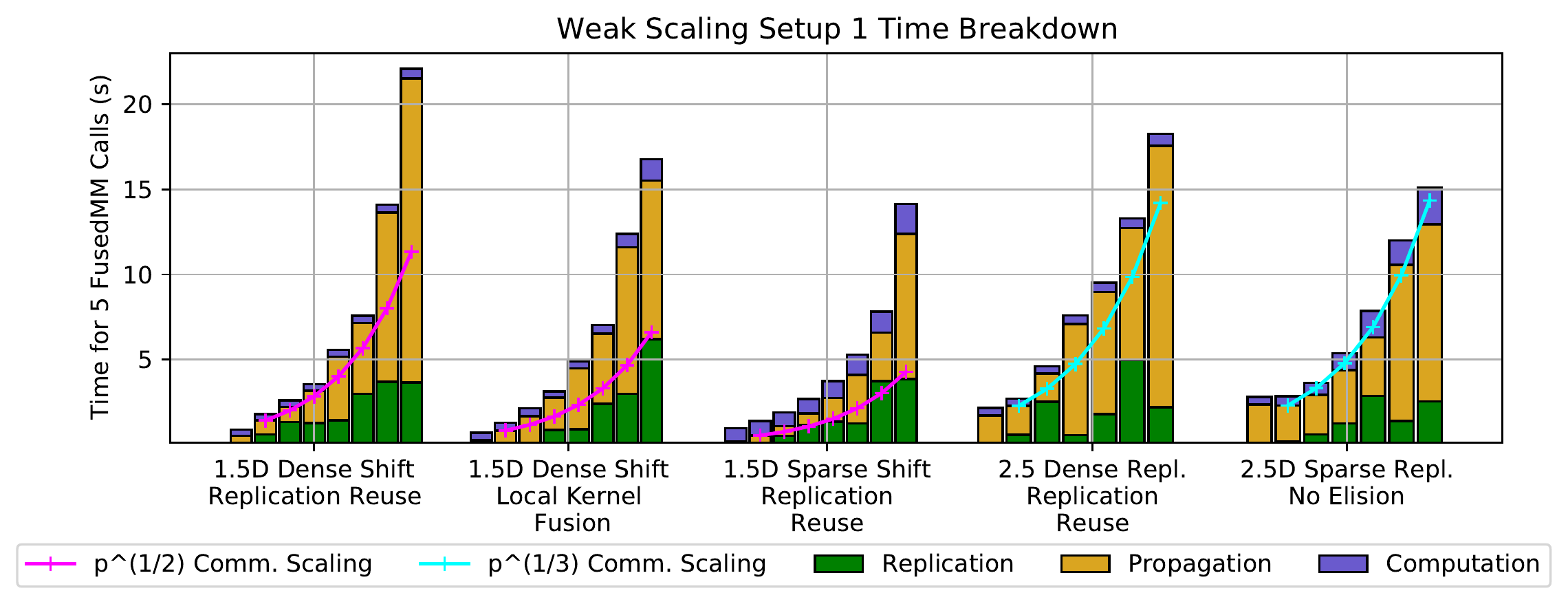}
  \vspace{0pt}
  \caption{Weak scaling time breakdown, setup 1. 
  Each section of bars indicates the time breakdown 
  for successively
  doubling processor counts, from $p=2$ on the left of
  each subsection to $p=256$ on the right.
  We expect
  $p^{1/2}$-communication scaling for 1.5D algorithms and
  $p^{1/3}$ for 2.5D algorithms.
  }
  \label{fig:weakscale_breakdown}
  \vspace{-12pt}
\end{figure}

While FLOPs per processor
remains constant from experiment to experiment, communication for our 1.5D algorithms scales 
as $O(n/\sqrt{p})$.
Doubling both $n$ and $p$ results in an expected increase in communication time of $\sqrt{2}$, 
giving a projected $\sqrt{p}$-scaling in communication time with processor count. Similarly, our 2.5D algorithms have
communication scaling of $O(n/\sqrt[3]{p^2})$, yielding a projected $\sqrt[3]{p}$ scaling in communication time
with processor count. Notice also under this strategy that $\phi$ remains constant at $32/256=1/8$, 
while the percentage of nonzeros in $S$ decays
exponentially. Figure \ref{fig:weakscale} (left) 
gives the results of these experiments, for the best observed replication factor at each processor count. Even though
each processor handles the same number of nonzeros across
experiments, the communication time (detailed in figure
\ref{fig:weakscale_breakdown}) quickly dominates the computation
time as we double the processor count. Among 1.5D 
algorithms, the
sparse shifting, dense replicating algorithm exhibits the best overall performance. We attribute this to 
the low, constant value of $\phi = \textrm{nnz}(S) / nr$ across the experiments. At 256 nodes, replication reuse allows the replication reusing sparse
shifting 1.5D algorithm to run 1.15 times faster than the 2.5D sparse replicating
algorithm, while the variant without communication elision runs 2\% slower than the 2.5D sparse
replicating algorithm. For 1.5D dense shifting algorithms,
both FusedMM elision strategies have roughly the same gain over the unoptimized back-to-back kernel sequence 
until the 256 node experiment. At 256 nodes, the 1.5D algorithm with local kernel fusion 
exhibits 1.38x speedup over its non-eliding counterpart, and 
replication reuse gives 1.16x speedup.

\noindent
\textbf{Setup 2:} Beginning with the same conditions for a single node as setup 1, 
node counts quadruple from experiment to experiment, as we both double
the side-length of the sparse matrix and double its nonzero count per row. The FLOPs per processor again remains
constant. Under this setup, however, the percentage of nonzeros of $S$ remains constant while $\textrm{nnz}(S) / nr$
doubles from experiment to experiment. Setup 2 provides insight into the scaling of the 1.5D sparse
shifting algorithm as the ratio $\phi$ successively doubles. Since the communication cost for
1.5D dense shifting algorithms does not depend on $\phi$ and scales as $O(n/\sqrt{p})$,
its communication cost should remain constant across experiments, while the
$O(n/\sqrt[3]{p^2})$ communication time scaling for 2.5D algorithms even implies a decrease in communication time. 
In practice, the decrease in node locality caused by scaling to high node counts renders
a decrease in overall time unlikely.

Figure \ref{fig:weakscale} (right half) shows the results, all of which take less than five seconds
due to better communication scaling compared to setup 1. Inverting the results from the 
first weak scaling benchmark, the 1.5D sparse shifting algorithm performs progressively worse as node count increases compared to the best
performer, the 1.5D dense shifting algorithm under local kernel fusion. At 256 nodes,
the 1.5D local kernel fusion algorithm is 1.94 times as fast as the sparse shifting
algorithm with replication reuse. For almost all cases, employing replication reuse 
or local kernel fusion results in nontrivial performance 
gains over an unoptimized sequence of SDDMM and SpMM.

\begin{figure}
    \centering
    \includegraphics[scale=0.48, trim={0cm 1cm 0cm 0cm, 
    clip}]{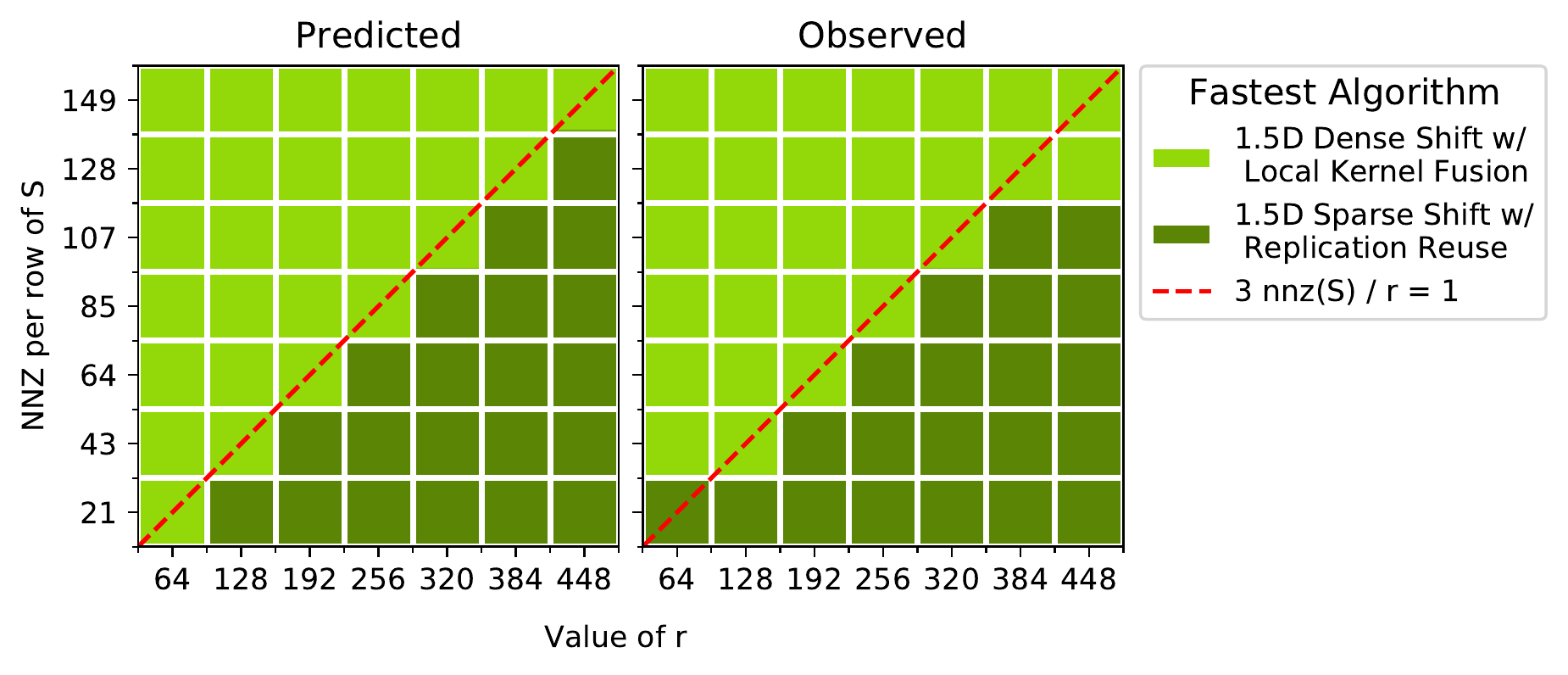}
    \vspace{1pt}
    \caption{Predicted, Observed Optimal Algorithms for $p=32, 
    m=2^{22}$. 
    2.5D algorithms were also benchmarked,
    but were neither predicted nor observed as best. The best observed replication factor
    in each configuration was used.}
   \label{fig:optimal_algorithm_selection}
   \vspace{-12pt}
\end{figure}

\subsection{Effect of Embedding Width $r$}

Compiled from 740 trials over different configurations, figure 
\ref{fig:optimal_algorithm_selection} shows the best algorithm out of the four that employ communication 
elision (along with the 2.5D sparse replicating algorithm) on a range of $r$-values and sparse matrix nonzero 
counts. 
As predicted, the 1.5D dense shifting algorithm performs better when $\phi = \textrm{nnz(S)} / nr$ is high,
while the 1.5D sparse shifting algorithm performs better for low $\phi$. Both 1.5D algorithms outperform the 2.5D
algorithms in theory as well as practice by margins comparable to those in figure \ref{fig:weakscale}. From figure 
\ref{fig:optimal_algorithm_selection}, we note that the optimal
algorithm choice is always a 1.5D sparse shifting or dense shifting algorithm
depending on the value of $\phi$, a conclusion that carries some caveats. 
Specifically, the performance of the 2.5D algorithms is hurt at
$p=32$ since the replication factor is constrained to be 
either 2 or 8. Our strong scaling experiments indicate that at high
enough node counts, 2.5D algorithms approach (and sometimes outperform) the
1.5D algorithms.

\begin{figure}
    \centering
    \includegraphics[scale=0.45, trim={0cm 1.4cm 0cm 0cm, 
    clip}]{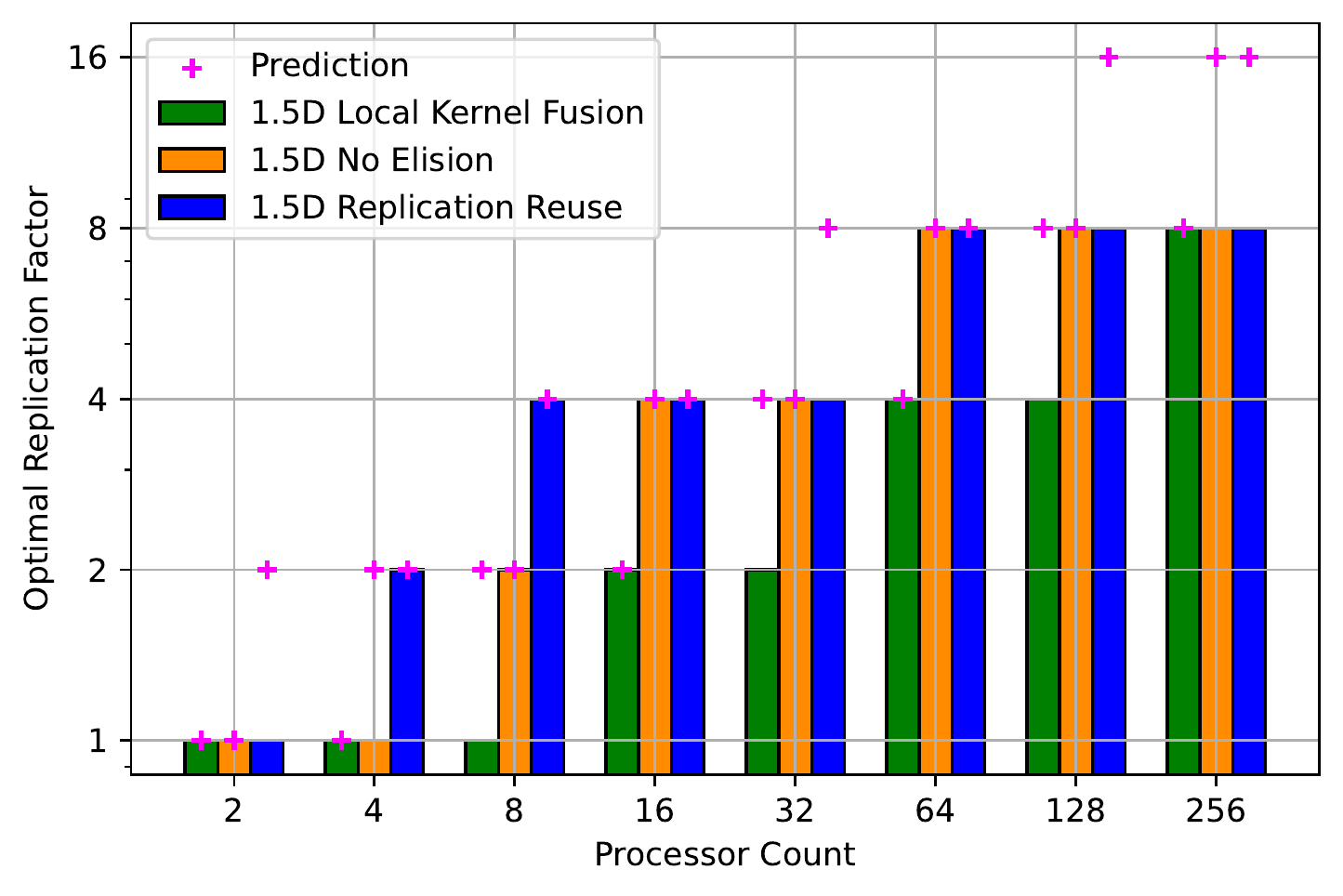}
    \vspace{5pt}
    \caption{Predicted, Observed Optimal Replication Factors for Weak Scaling Experiments.}
    \label{fig:replication_factor_effect}
\end{figure}

For weak scaling setup 1, figure \ref{fig:replication_factor_effect} gives the 
predicted and observed optimal replication factor as a function of the processor count for 1.5D dense shifting algorithms. As predicted, the 
optimal replication factor $c$ for the 1.5D algorithm with replication 
reuse is at least the optimal replication factor for the unfused 
algorithm. The latter, in turn, is at least the optimal replication factor of the local kernel fusion
algorithm. The plot experimentally confirms that our fused algorithms save communication 
by changing the optimal replication factor in addition to decreasing the 
number of communication rounds.

Our predictions of optimal replication factor match 
the observed optimal values on most experiments. When they disagree,
our theory tends to overestimate for two reasons: first, 
we did not test replication factors higher than 8 for our weak scaling
experiments due to memory constraints, leading to the gap at the right end of the figure. 
Second, we used an ordering of MPI ranks that maximized locality 
within each layer of the processor grid. As a result,
communication within the fiber axis (i.e. replication costs) are more expensive due to lack of node locality, a fact that we verified by comparing against 
an MPI rank order that optimized for locality along each fiber.

\subsection{Strong Scaling on Real-World Matrices}

\begin{figure*}
    \centering
  \includegraphics[clip, trim={0cm, 0cm, 0cm, 0cm}, scale=0.45]{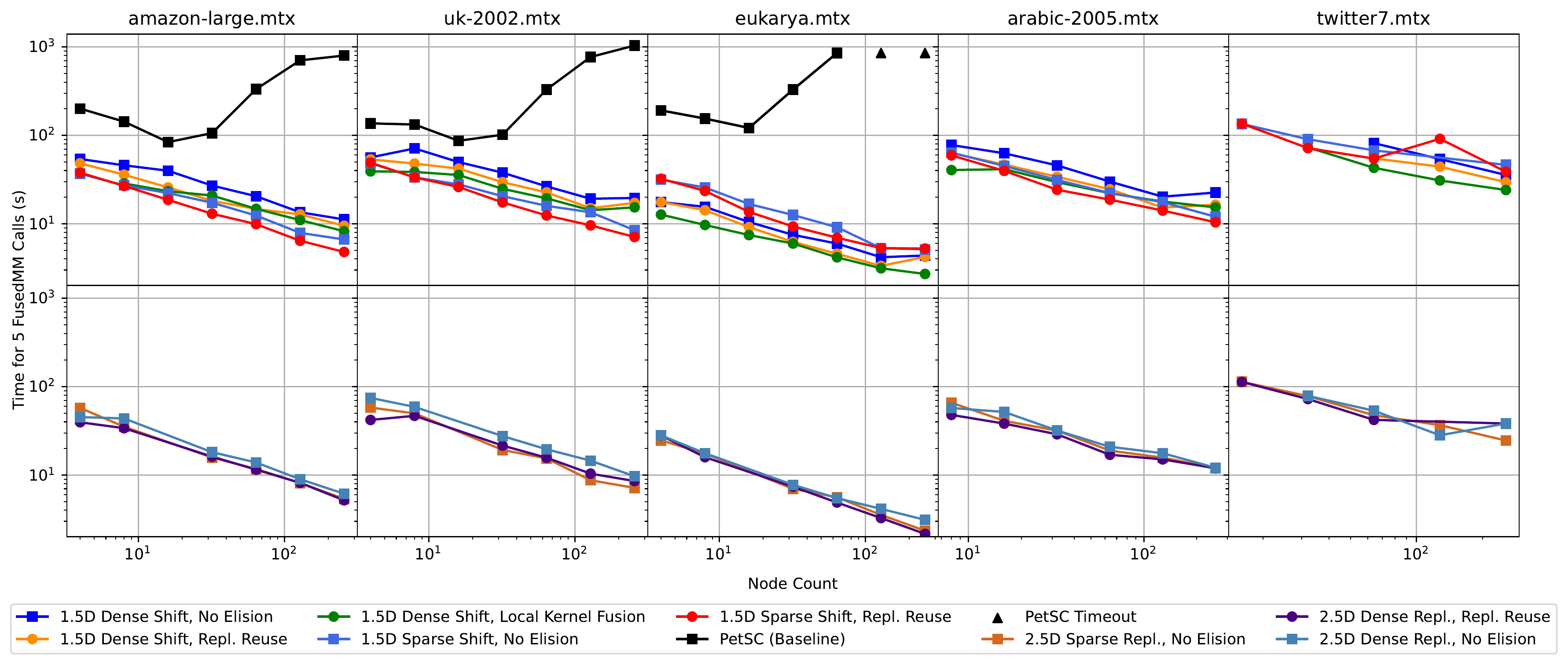}
  \vspace{-7pt}
  \caption{Strong scaling experiments,  $r=128$.
  We benchmarked our algorithms on 5 FusedMM calls and the baseline implementation \petsc\ with
  10 SpMM calls. Black triangles indicate that \petsc\ took longer than three hours to complete
  the trial. We benchmarked with a minimum of 4 nodes for the smaller workloads, 8 nodes
  for Arabic-2005, and 16 nodes for twitter7.
  }
  \label{fig:strongscale}
  \vspace{-12pt}
\end{figure*}

We conduct strong scaling experiments on up to 256 KNL nodes with $r = 128$ on five real-world matrices 
given in table \ref{table:strongscale_matrices} containing up to $\approx 1.5$ billion nonzeros. amazon-large.mtx, uk-2002.mtx, arabic-2005.mtx, and twitter7.mtx were taken from the Suitesparse matrix collection
\cite{suitesparse}, while eukarya.mtx contains protein sequence alignment
information for eukaryotic genomes \cite{hipmcl}. With $\approx 360$ million nonzeros but 
only $\approx 3$ million vertices, eukarya is the most dense at 111 
nonzeros per row, compared with roughly 16 nonzeros per row for both uk-2002 and amazon-large. Twitter7 and Arabic-2005 fall between the two 
extremes at 28-35 nonzeros per row, but have significantly more total nonzeros compared to 
the other matrices.
We expect that the 1.5D sparse shifting and 2.5D sparse replicating algorithms 
will exhibit better communication performance on the sparse Amazon matrix, while
the 1.5D dense shifting and 2.5D dense replicating algorithms become optimal for eukarya and
and twitter7. Because we choose
$r = 128$ conservatively to avoid allocating large amounts of RAM at small node counts, we enforce a minimum
replication factor of 2 for the 1.5D sparse shifting algorithm at 256 nodes (since we cannot divide 128 into
more than 256 parts for $c = 1$).

\begin{table}[]
\centering
\begin{tabular}{|c|c|c|c|}
\hline
\textbf{Matrix}  & \textbf{Rows} & \textbf{Columns} & \textbf{Nonzeros} \\ \hline
amazon-large.mtx & 14,249,639      & 14,249,639         & 
230,788,269 \\
uk-2002.mtx      & 18,484,117      & 18,484,117         & 298,113,762         \\
eukarya.mtx      & 3,243,106       & 3,243,106          & 359,744,161         \\
arabic-2005.mtx  & 22,744,080      & 22,744,080         & 639,999,458         \\
twitter7.mtx     & 41,652,230      & 41,652,230         & 1,468,365,182       \\\hline
\end{tabular}
\vspace{2mm}
\caption{Matrices Used in Strong Scaling Experiments}
\label{table:strongscale_matrices}
\vspace{-22pt}
\end{table}

Figure \ref{fig:strongscale} gives the results of the strong scaling experiments; the performance of each algorithm 
is its best runtime over replication factors from 1 through 16. As we expect, the 1.5D sparse
shifting algorithm with replication reuse performs best on the amazon-large and uk-2002 matrices, 
while it is the second worst on eukarya. 
With 256 nodes on uk-2002, the 1.5D sparse shifting algorithm with replication 
reuse performs 1.19x faster than the version without communication elision, 
and it performs 2.1x faster than the dense shifting fused algorithm with 
local kernel fusion. On eukarya with 256 nodes, the dense shifting 
algorithm with local kernel fusion performs 1.6x faster than 
the version without communication elision and 1.9x faster than
the sparse shifting algorithm. Among 2.5D algorithms, 
the dense replicating algorithm with replication reuse and the sparse replicating algorithm have
similar performance, and both outperform 
the 2.5D dense replicating algorithm with no communication elision.
As predicted by our theory, the dense replicating algorithm has slightly better performance on eukarya.mtx at high node counts, even outperforming all of our 1.5D algorithms.

\subsection{Applications}
Here, we plug in our distributed memory algorithms to machine learning
applications to benchmark their performance. We focus on two: collaborative
filtering with alternating least squares, and graph neural networks with
self-attention.

\textbf{Collaborative Filtering with ALS: } Collaborative filtering 
attempts to factor a matrix $C \in \RR^{m \times n}$ 
as $C = A B^T$, for $A \in \RR^{m \times r}, B \in \RR^{n \times r}$; however,
we only have access to a set of sparse observations of $C$, denoted as 
the sparse matrix $\tilde C$ with nonzero indicators $S$. 
We iteratively minimize the loss, which is the Frobenius norm of  
$\tilde C - \textrm{SDDMM}(A, B, S)$. The SDDMM kernel in the loss 
also appears (along with SpMM) in the iterative update equations. 

The ALS method 
alternately keeps
$A$ or $B$ fixed and, for each row $x$ of the matrix to optimize for, solves a least squares
problem of the form $Mx = b$. 
These least squares problems are distinct for each row 
due to the varying placement of nonzeros within each column of $S$. If we use Conjugate Gradients (CG) as the 
least squares solver, Zhao and Canny \cite{canny_big_nodate} exhibit the technique 
of batching computation of the query vectors 
$Mx$ for all rows at once using a FusedMM operation. 

2.5D sparse replicating and 2.5D dense replicating algorithms suffer slight penalties for this application compared to 
1.5D algorithms, as the output distributions of the dense matrices are shifted and transposed, respectively, 
compared to the input distributions. Since
the output query vectors become (after some additional manipulation) inputs to the next CG iteration, 2.5D algorithms must pay
to shift the input and output distributions at each step. We 
benchmark 20
CG iterations with our distributed algorithms embedded: 10 to
optimize the matrix $A$, and 10 to optimize the matrix $B$.

\textbf{Graph Attention Network (Forward Pass Workload)}: 
Consider a graph with adjacency
matrix $S \in \set{0, 1}^{n \times n}$. Conventional GNNs contain a series of layers, 
with an $r$-length vector of features associated with each node as the input to a layer. 
These node embeddings are held in an $n \times r$ matrix $A$, and the GNN layer applies a small linear transformation
$W \in \RR^{r \times r'}$ to the embedding matrix before performing a convolution to sum the feature 
vectors at neighbors of any node $x$ as to become the new feature vector of $x$. 
Application of a nonlinear function $\sigma$, typically follows the convolution, 
giving the final layer output 
$A' \in \RR^{n \times r'}$ as $A' = \sigma (SAW)$.

The subsequent GNN layer takes $A'$ as an input, with the final layer generating a node-level or graph-level prediction.
Graph attention networks (GATs) modify this architecture by weighting edges with a self-attention score computed using the embeddings
of the incident nodes. A single self-attention head \cite{velickovic_graph_2018} replaces $S$ with $S' = \sigma_{\textrm{LeakyReLU}}(S * (A \cdot_{\mathit{GAT}} A^T))$.
Here, $A \cdot_{\mathit{GAT}} A^T$ is an $n \times n$ matrix with entries given by 
$(A \cdot_{\mathit{GAT}} A^T)_{ij} = a^T (A_{i:} \vert\vert A_{j:})$, where $\vert\vert$ denotes concatenation and $a^T$ is a trainable vector. The computation of $S'$ involves a slight 
modification of Eq.~\ref{eq:sddmm} and has an identical communication pattern to SDDMM. A multi-head GAT concatenates the outputs of several attention heads with distinct trainable weight matrices $W$ and weight vectors $a$. We simulate the forward pass workload of this multi-head GAT architecture using random weight matrices to focus on the communication reduction and scaling of our distributed-memory algorithms. 

\begin{figure}
    \centering
    \includegraphics[scale=0.45, trim={0cm 0.2cm 
    0cm 0cm}, clip]{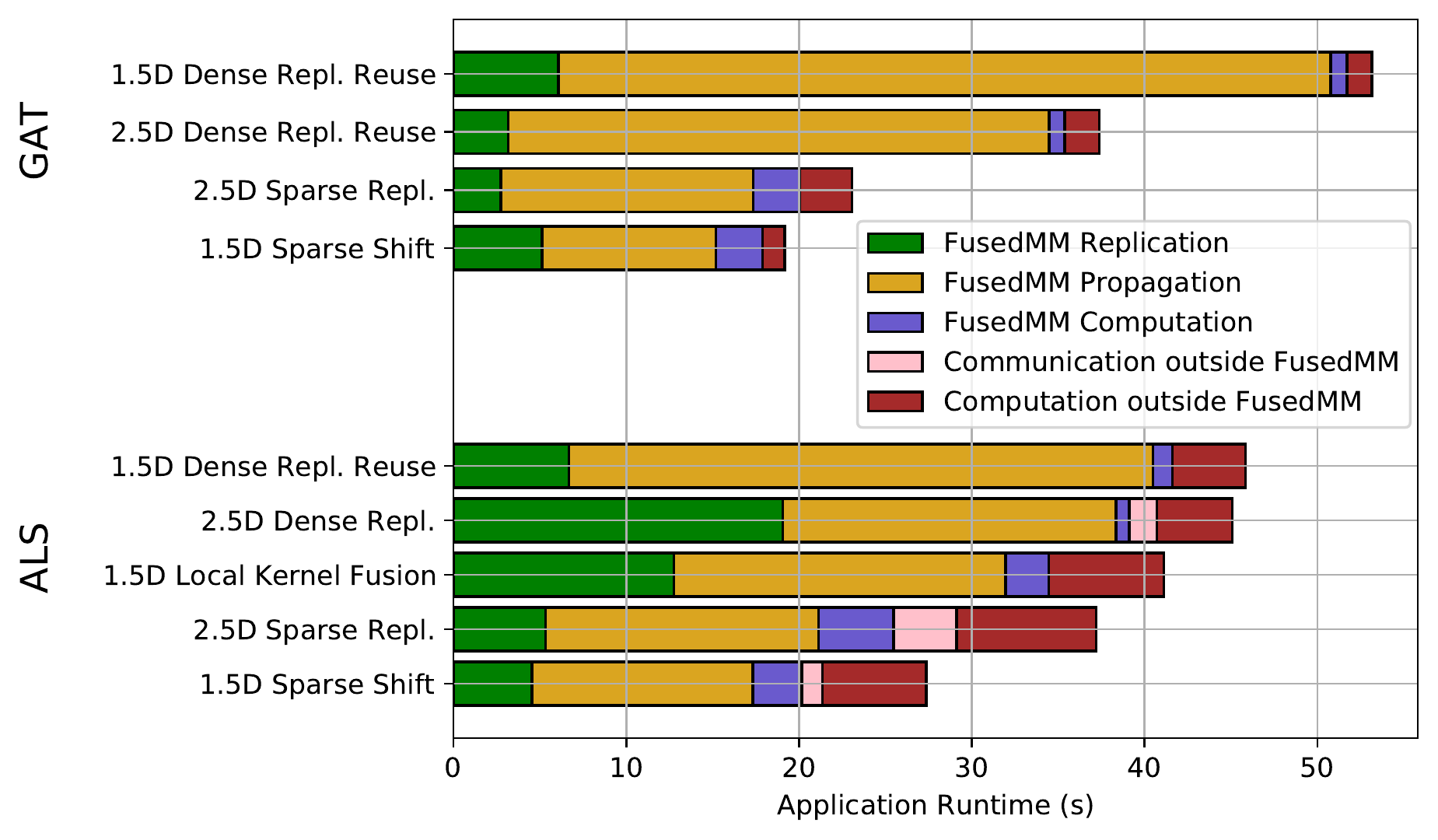}
    \caption{Alternating Least Squares and Graph Attention Network Forward Pass
    Performance on Amazon.mtx. We use 256 nodes and $r=128$. The 1.5D algorithm
    with local kernel fusion was not benchmarked for GATs, as it is incompatible
    with softmax regularization of learned edge weights.}
    \label{fig:app_benchmark}
    \vspace{-12pt}
\end{figure}

Figure~\ref{fig:app_benchmark} shows the time breakdown of our applications,
both inside the FusedMM kernels and in the rest of the application. 
The ALS application exhibits some variation in communication and computation time spent
outside FusedMM. The variation is only partially due to additional processor to processor communication to compute distributed dot products, which is higher for the sparse
replicating and shifting algorithms. More significantly, at the high replication 
factors (8 and 16) used in the experiment, the row-major local dense matrices are 
extremely tall-skinny for the 1.5D sparse
shifting and 2.5D sparse replicating algorithms compared to the other variants. Careful analysis 
of the CG solver revealed that the dense batch dot product operation 
requires a long sequence of poorly performing 
dot product calls on short vectors. Since
the sequence grows linearly with local matrix height, hundreds of thousands 
of these calls slow performance for the 
1.5D sparse shifting and 2.5D sparse replicating algorithms. Calling an optimized
batched BLAS library would fix this issue, which we leave as future work.

\section{Conclusions and Further Work}
We gave a theoretical communication analysis of distributed memory sparsity 
agnostic approaches for SDDMM and FusedMM. Our theory predicted 
communication savings using two distinct approaches to combine the two kernels, and we
observed those benefits in our strong and weak scaling experiments. In both
theory and practice, 1.5D sparse shifting and 2.5D sparse replicating 
algorithms perform better when $S$ has far fewer nonzeros 
compared to either dense matrix. When $S$ has a higher nonzero count, 
dense shifting 1.5D and dense replicating 2.5D algorithms win out. 

Further performance improvement may be possible
by overlapping communication in the propagation phase of any of our algorithms 
with local computation. Such an implementation might require one-sided MPI or a similar 
protocol for remote direct memory access (RDMA) without CPU involvement.

\section*{Acknowledgements}
This material is based upon work supported in part by the U.S. Department of Energy, Office of Science, 
Office of Advanced Scientific Computing Research, Department of Energy Computational Science
Graduate Fellowship under Award Number DE-SC0022158. This work is also supported by the Office of Science of the DOE under contract number DE-AC02-05CH11231. 

We used resources of the NERSC supported by the Office of Science of the DOE under Contract No. DE-AC02-05CH11231. 

\section*{Disclaimer}
This report was prepared as an account of work sponsored by an agency of the United
States Government. Neither the United States Government nor any agency thereof, nor any of their
employees, makes any warranty, express or implied, or assumes any legal liability or responsibility for the
accuracy, completeness, or usefulness of any information, apparatus, product, or process disclosed, or
represents that its use would not infringe privately owned rights. Reference herein to any specific
commercial product, process, or service by trade name, trademark, manufacturer, or otherwise does not
necessarily constitute or imply its endorsement, recommendation, or favoring by the United States
Government or any agency thereof. The views and opinions of authors expressed herein do not
necessarily state or reflect those of the United States Government or any agency thereof.

\bibliographystyle{IEEEtran}
\bibliography{main}

\end{document}

%% file: preamble.tex
\usepackage{fancyhdr}

\usepackage{amsmath,amssymb, amsthm}
\usepackage{graphicx}
\usepackage{accents}
\usepackage{hyperref}
\usepackage{bbm}
\usepackage{bm}
\usepackage[ruled,vlined, linesnumbered]{algorithm2e}
\usepackage{booktabs}
\usepackage{float}
\usepackage{multirow}

\newcommand{\RR}{\mathbb{R}}

\newcommand{\paren}[1]{\left( #1 \right)}

\newcommand{\set}[1]{\{#1\}} 
\newcommand{\gen}[1]{\langle #1 \rangle}

\newcommand{\erdosrenyi}{Erd\H{o}s-R\'{e}nyi}
\newcommand{\petsc}{PETSc}